\documentclass[sigconf]{acmart}

\usepackage{balance}

\def\BibTeX{{\rm B\kern-.05em{\sc i\kern-.025em b}\kern-.08emT\kern-.1667em\lower.7ex\hbox{E}\kern-.125emX}}

\usepackage{amsmath}
\usepackage{amsfonts}
\usepackage{bm}
\usepackage{color}
\usepackage{graphicx}
\usepackage[noadjust]{cite}
\usepackage{url}
\usepackage{setspace}
\usepackage{subcaption}
\usepackage{booktabs}
\usepackage{mathtools}
\usepackage{multirow}
\usepackage{float}
\usepackage{caption} 
\usepackage{algorithm}
\usepackage{algorithmic}
\usepackage{cleveref}
\usepackage{makecell}
\usepackage{xcolor,colortbl}

\usepackage{tabularx}
\definecolor{lightlightgray}{gray}{0.85}
\definecolor{lllgray}{gray}{0.9}
\definecolor{llllgray}{gray}{0.95}
\crefname{equation}{Eq.}{Eq.}
\crefname{section}{Section}{Sections}
\crefname{subsection}{Section}{Sections}
\crefname{subsubsection}{Section}{Sections}
\crefname{figure}{Figure}{Figures}
\crefname{table}{Table}{Tables}
\crefname{subfigure}{Figure}{Figures}
\crefname{algocf}{Algorithm}{Algorithms}

\usepackage{enumitem}
\setlist{nosep,after=\vspace{0.0\baselineskip},leftmargin=10pt}
\setlist[itemize]{leftmargin=0.8\parindent,listparindent=\parindent,parsep=0.2\parskip,itemsep=0.02in,topsep=0.02in,after=\vspace{0in}}
\newcommand{\xhdr}[1]{\vspace{0.06in}\noindent{{\bf #1}}}
\newcommand{\xihdr}[1]{\vspace{0.03in}\noindent{{\bf #1}}}
\newcommand{\dataname}[1]{\textbf{\textit{#1}}}
\AtBeginDocument{%
  \providecommand\BibTeX{{%
    \normalfont B\kern-0.5em{\scshape i\kern-0.25em b}\kern-0.8em\TeX}}}

\copyrightyear{2020}
\acmYear{2020}
\setcopyright{acmcopyright}
\acmConference[WSDM '20]{The Thirteenth ACM International Conference on Web Search and Data Mining}{February 3--7, 2020}{Houston, TX, USA}
\acmBooktitle{The Thirteenth ACM International Conference on Web Search and Data Mining (WSDM '20), February 3--7, 2020, Houston, TX, USA}
\acmPrice{15.00}
\acmDOI{10.1145/3336191.3371855}
\acmISBN{978-1-4503-6822-3/20/02}

\begin{document}

\title{Addressing Marketing Bias in Product Recommendations}

\settopmatter{authorsperrow=4}
\author{Mengting Wan}
\affiliation{%
 \institution{Airbnb, Inc.}}
\email{mengting.wan@airbnb.com}
\authornote{Work was done while the author was at UC San Diego.}

\author{Jianmo Ni}
\affiliation{%
  \institution{UC San Diego}}
\email{jin018@eng.ucsd.edu}

\author{Rishabh Misra}
\affiliation{%
  \institution{Twitter, Inc.}}
\email{rmisra@twitter.com}

\author{Julian McAuley}
\affiliation{%
  \institution{UC San Diego}}
\email{jmcauley@eng.ucsd.edu}

\fancyhead{}


\begin{abstract}
Modern collaborative filtering algorithms seek to provide personalized product recommendations by uncovering patterns in consumer-product interactions. 
However, these interactions can be biased by how the product is marketed, for example due to the selection of a particular human model in a product image. These correlations may result in the underrepresentation of particular niche markets in the interaction data; for example, a female user who would potentially like motorcycle products may be less likely to interact with them if they are promoted using stereotypically `male' images.

In this paper, we first investigate this correlation between users' interaction feedback and products' marketing images on two real-world e-commerce datasets. We further examine the response of several standard collaborative filtering algorithms to the distribution of consumer-product market segments in the input interaction data, revealing that marketing strategy can be a source of bias for modern recommender systems. In order to protect recommendation performance on underrepresented market segments, we develop a framework to address this potential marketing bias. Quantitative results demonstrate that the proposed approach significantly improves the recommendation fairness across different market segments, with a negligible loss (or better) recommendation accuracy.

\end{abstract} 

\maketitle

\section{Introduction}
By connecting users to relevant products across the vast range available on e-commerce platforms, modern recommender systems are already ubiquitous and critical on
both sides of the market, i.e.,~consumers and product sellers. Among recommendation algorithms used in practice, many fall under the umbrella of \textit{collaborative filtering} \citep{sarwar2001item,linden2003amazon,herlocker1999algorithmic,koren2009matrix}, which collect and generalize users' preference patterns from logged consumer-product interactions (e.g.~purchases, ratings). 
These feedback interactions can  be \emph{biased} by multiple factors, potentially 
surfacing
unfair (or irrelevant) recommendations 
to users or items underrepresented in the input data. Such phenomena have already raised some attention from the recommender system community: a handful of types of algorithmic biases have been addressed, including selection bias \citep{schnabel2016recommendations}, popularity bias \citep{yang2018unbiased}, and several fairness-aware recommendation algorithms have been proposed \citep{beutel2019fairness,burke2018balanced}. In this paper, we focus on a relatively underexplored factor---\textit{marketing bias}---in consumer-product interaction data, and study how recommendation algorithms respond to its effect.

\begin{figure}[t!]
	\centering
	\includegraphics[width=\linewidth]{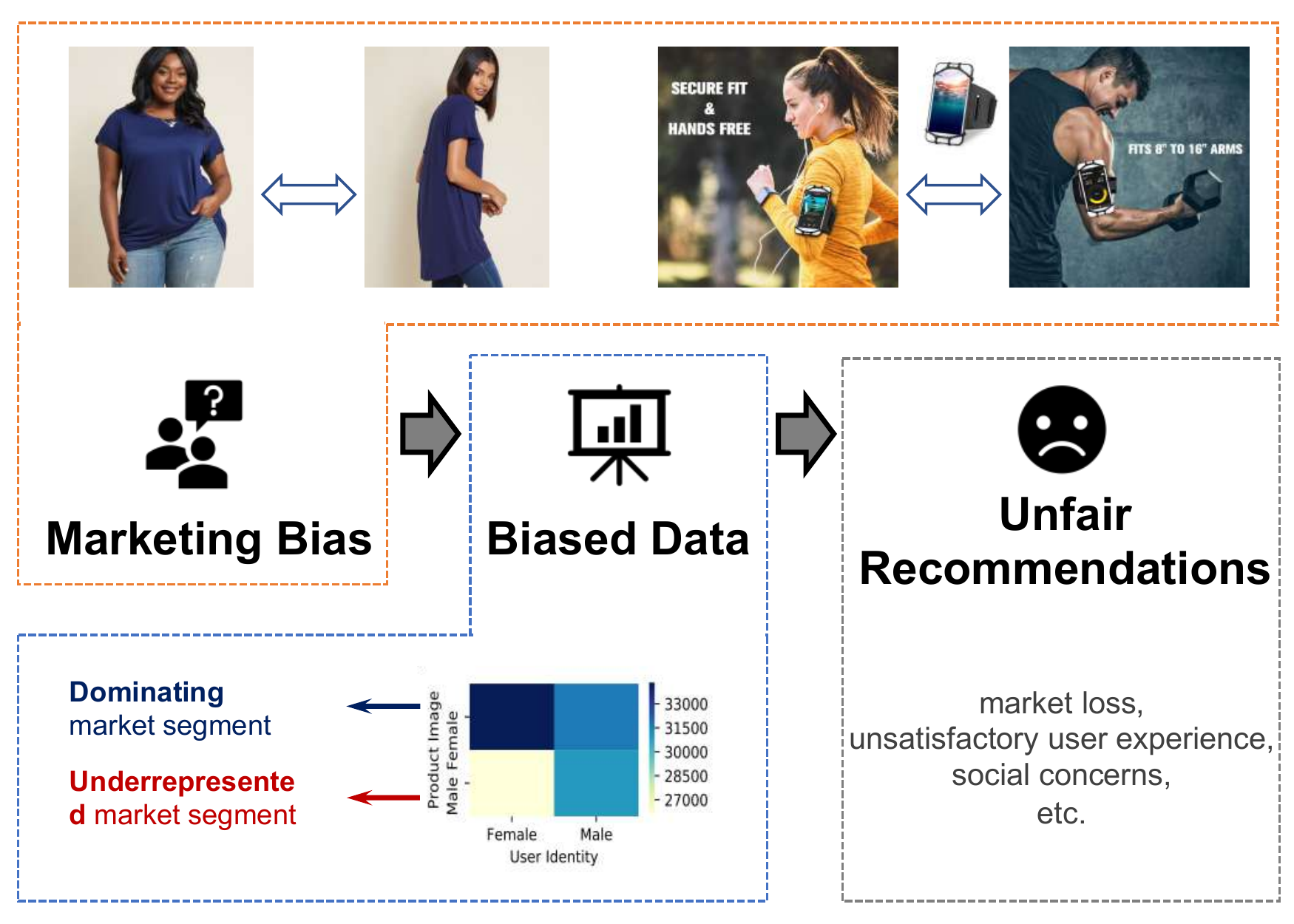}
	\caption{Two illustrative examples on how the same product can be marketed using different human images (different body shapes, different genders). These marketing strategies could affect consumers' behavior thus resulting in a biased interaction dataset, which is commonly used as the input for modern recommender systems.}\label{fig:illustration}
\end{figure}

We are particularly interested in the human factors, such as the profile of the human model in a product image,  reflected in a product's marketing strategies, which (as indicated in previous marketing studies) could possibly affect consumers' interactions and satisfaction
\citep{grubb1967consumer,grubb1968perception,birdwell1968study}. A common hypothesis (known as `self-congruence') is that a consumer may tend to buy a product because its public impression (in our case a \textit{product image}), among other alternatives, is \textit{consistent} with one's self-perceptions 
(\textit{user identity}) \citep{grubb1968perception}. Based on this assumption, the selection of human models for a product (as shown in \cref{fig:illustration}, a product can be represented by models with different body shapes or different genders) could influence a consumer's behavior. For example, a female user may be less likely to interact with an armband product which is presumably gender-neutral but marketed 
exclusively via `male' images.
As with many other types of bias, this could 
lead to
underrepresentation of some niche market segments in the input data for a recommender system.
Note if undesired patterns are propagated into recommendation results (e.g.~even fewer male-represented products are recommended to the potential female users), utility from both sides of the marketplace could be harmed. That is, product retailers may lose potential consumers while users may be struggling 
to find
relevant products. As a consequence, serious ethical and social concerns could be raised as well. 

In this work, we seek to understand 1) if such a marketing bias exists in real-world e-commerce datasets; 2) how common collaborative filtering algorithms interact with these potentially biased datasets; and 3) how to alleviate such algorithmic bias (if any) and improve the market fairness of recommendations.
We conclude our contributions as follows.
\begin{itemize}
    \item We collect and process two e-commerce datasets from ModCloth and Amazon.
    Then we conduct an observational study to investigate the relationship between interaction feedback and product images (reflected in the selection of a human model) as well as user identities. Different types of correlations in varying degrees can be observed in these two datasets.
    \item We implement several common collaborative filtering algorithms and study their responses to the above patterns in the input data. For most algorithms, we find 1) systematic deviations across different consumer-product segments in terms of rating prediction error, and 2) notable deviations of the resulting recommendation outputs from the real interaction data.
    \item Note that as the marketing bias could be intricately entangled with users' intrinsic preferences, our goal in this work is not to pursue the absolute parity of recommendations (e.g.~keep recommending products represented by human images which were constantly unfavored by a user). Rather, we expect a fair algorithm is supposed \textit{not to worsen the market imbalance} in interactions.
    We thus propose a fairness-aware framework to address it by calibrating the parity of prediction \textit{errors} across different market segments. 
    Quantitative results indicate that our framework significantly improves recommendation fairness and provides better accuracy-fairness trade-off against several baselines.
\end{itemize}

\section{Related Work}
This work is partially motivated by the well-known `self-congruity' theory in marketing research, which is defined as the match between the product/brand image and the consumer's true identity and the perception about oneself \citep{grubb1967consumer,grubb1968perception,birdwell1968study}. Many previous marketing studies focus on assessing this theory by quantifying and validating it through statistical analysis on a small amount purchase transaction data or the feedback in questionnaires \citep{sirgy1982self,sirgy1997assessing,malhotra1988self,kressmann2006direct}. Following self-congruity theory, products can be advertised in a way to match their target consumers' images thus establishing product stereotypes \citep{grau2016gender}. Our work is distinguished with these studies from a more computational perspective, by identifying and studying the potential marketing bias for recommender systems on large-scale e-commerce interaction datasets.

Our analysis is related to previous work which examines particular types of biases in real-world interactions and their effects in recommendation algorithms, including the popularity effect and catalog coverage \citep{jannach2015recommenders}, the bias regarding the book author gender for book recommenders \citep{ekstrand2018exploring}, and the herding effect in product ratings \citep{zhang2017modeling}. 

Another closely related line of work includes developing evaluation metrics and algorithms to address fairness issues in recommendations. `Unbiased' recommender systems with missing-not-at-random training data are developed by considering the propensity of each item \citep{schnabel2016recommendations,joachims2017unbiased}. A fairness-aware tensor-based algorithm is proposed to address the absolute statistical parity (i.e.,~items are expected to be presented at the same rate across groups) \citep{zhu2018fairness}. Several fairness metrics and their corresponding algorithms are proposed for both pointwise prediction frameworks \citep{burke2018balanced,yao2017beyond} and pairwise ranking frameworks \citep{beutel2019fairness}. Methodologically, these algorithms can be summarized as reweighting schemes where underrepresented samples are upweighted \citep{schnabel2016recommendations,joachims2017unbiased,burke2018balanced} or 
schemes where additional fairness terms are added to regularize the model \citep{yao2017beyond,beutel2019fairness,abdollahpouri2017controlling}. 

Note that most of the above studies focus on bias and fairness on one side of the market only (i.e.,~either user or producer). Our concern about marketing bias is that it could affect fairness for both consumers and product providers. Without global market fairness in mind, the imbalance of the consumer-product segment distribution could be exacerbated through the deployment of recommendation algorithms.
Multi-sided fairness is addressed by Burke \textit{et al} \citep{beutel2019fairness} by considering C(onsumer)-fairness and P(rovider)-fairness. Trade-off between accuracy and fairness in two-sided marketplaces is further explored and a counterfactual framework is proposed to evaluate different recommendation policies without extensive A/B tests \citep{mehrotra2018towards}.
However the CP-fairness condition where fairness is protected for both sides \textit{at the same time} still remains an open question.

\section{Data Collection and Preprocessing}
We introduce two real-world e-commerce datasets collected from a women's clothing website \textit{ModCloth}\footnote{\url{https://www.modcloth.com/}} and the \textit{Electronics} category on \textit{Amazon}.\footnote{\url{https://www.amazon.com/}} These datasets enable us to study the marketing bias induced by the selection of a human model
with respect to \textit{body shape} for clothing products, and investigate the effects from the \textit{gender} of human models for electronics products.
Detailed information about these datasets can be found in \cref{tab:stats}.
Datasets in this paper are available at \url{https://github.com/MengtingWan/marketBias}. 

\begin{table}[]
    \centering
    \begin{tabular}{lrr}
\toprule
 &                                      \textbf{ModCloth} &                                        \textbf{Electronics} \\
\midrule
\#review    &                                         99,893 &                                            1,292,954 \\
\#item      &                                          1,020 &                                               9,560 \\
\#user      &                                         44,783 &                                            1,157,633 \\
time span & 2010-2019 & 1999-2018 \\
\midrule
bias type & body shape & gender \\
\hline
product image  & \makecell[r]{Small (838)\\ Small\&Large (182)} &  \makecell[r]{Female (4,090)\\ Female\&Male (2,466)\\ Male (3,004)} \\
\hline
user identity &  \makecell[r]{Large (9,395)\\ Small (30,140)\\ N/A (5,248)} &   \makecell[r]{Female (71,043)\\ Male (61,350)\\ N/A (1,025,240)} \\
\bottomrule
\end{tabular}
    \caption{Basic statistics of the \dataname{ModCloth} and \dataname{Electronics} datasets.}\label{tab:stats}
\end{table}

Note that our datasets are not perfect, e.g.~errors and selection bias can be introduced via scraping, parsing and processing, control of several confounding factors including inventory status,
(etc.).
Our intention here is neither to make any normative claims regarding the distributions in the above two applications, nor draw any causal conclusions. Rather, we simply describe the current state of these datasets and \emph{study how recommendation algorithms interact with these 
data.}

\subsection{ModCloth}
\dataname{ModCloth} is an e-commerce website which sells women's clothing and accessories. One 
unique property of this data is that many products include two human models with different body shapes (as shown in \cref{fig:illustration}) and measurements of these models. In addition, users can optionally provide the product sizes they purchased and fit feedback (`Just Right', `Slightly Larger', `Larger', `Slightly Smaller' or `Smaller') along with their reviews. Therefore we focus on the dimension of \textit{human body shape} as the source of marketing bias in this dataset.

\xhdr{Product Image Group (Body Shape).}
We start with the clothing products included in an existing public dataset \citep{misra2018decomposing}, re-scrape their landing pages, collect related model size measurements and all review ratings. We normalize their product sizes as `XS', `S', `M', `L', `XL', `1X', `2X', `3X' and `4X' according to the provided size charts.\footnote{e.g.~\url{https://www.modcloth.com/size-guide.html}} 
Products with only one human model wearing a relatively small size (`XS', `S', `M' or `L') are labeled as the `Small' group while products with two models (an additional model wearing a plus-size: 1X', `2X', `3X' or `4X') are referred as the `Small\&Large' group. 

\xhdr{User Identity Group (Body Shape).}
We then calculate the average size each user purchased and classify users into `Small' and `Large' groups based on the same standard as the product body shape image.

We observe that all products offer the complete spectrum of sizes, while 70\% of these products are interacted with by at least one user from the `Large' group and 97\% are interacted with by the `Small' group.
Thus we conclude that most users are able to consume most products at some point within the time frame of our dataset.

Ultimately we collect nearly 100K reviews about 1,020 clothing products from 44,783 users, where around 90\% of users can be matched to the above identity groups.

\subsection{Electronics}
\dataname{Electronics} is another review dataset collected from the \textit{Electronics} category on Amazon with \textit{Clothing} as an auxiliary category. This dataset is built on top of the public \textit{Amazon 2018 Dataset} \citep{ni2019justifying} and further processed to facilitate the research goals in this paper.
We regard the \textit{gender} as the target marketing bias on this dataset.

\xhdr{Product Image Group (Gender).}
In the \textit{Amazon 2018 Dataset}, we keep all pictures associated with electronic products\footnote{All products attached to the `Men' or `Women' categories are removed.} and run human model detection through an industrial body/face detection API provided by Face++.\footnote{\url{https://www.faceplusplus.com/}} 
The results include whether any human bodies/faces are included in the pictures, as well as gender predictions of these detected models. We only keep products where human models are detected in their associated pictures and treat
them as three types of product gender image based on the selection of these human models: `Female' (only female models are included), `Male' (only male models are included) and `Female \& Male' (both female and male models are detected, not necessarily in the same picture).

We then involve 3 human labelers to conduct
validations on this dataset, where label conflicts are resolved by majority voting. 3,000 randomly sampled pictures are manually labeled regarding
1) if they notably include human models; 2) the gender image from `Female Exclusive', `Male Exclusive' or `Both Female \& Male' (if multiple models are included in a single picture). We evaluate the human model detection results from the API based on these labels and find a high precision (96\%) regarding the human model detection but a relatively low recall (53\%). Note in our setting we are happy to discard ambiguous cases (sacrifice some recall) for the sake of high precision.
We later randomly sample 100 products and manually decide if these products preserve any gender constraints based on their descriptions. Although 4 out of 100 products exhibit gender implications,\footnote{e.g.~\url{https://www.amazon.com/gp/product/B00HX19EDI}} we don't find any strict constraints which prevent the unfavorable user identity group from consuming these products.

\begin{table}
    \centering
    \setlength{\tabcolsep}{5pt}
    \begin{tabular}{p{0.8in}p{2.3in}}
    \toprule
      \textbf{Term/Symbol}   & \textbf{Description} \\
     \midrule
      product image   & the public impression of a product; attributes of the human models included in the product pictures are used in this work, e.g.~body shape, gender\\[0.5mm]
      user identity & the perception of oneself; we use the same dimension of attribute as in product image \\[0.5mm]
      $m$, $n$ & user identity group, product image group, e.g.~female/male\\[0.5mm]
      $M$, $N$ & the number of possible user identity groups and product image groups\\[0.5mm]
      $s_{u,i}$, $r_{u,i}$, $e_{u,i}$ & predicted user $u$'s preference score on product $i$, user $u$'s rating score on product $i$, prediction error $e_{u,i} = s_{u,i} - r_{u,i}$\\[0.5mm]
      $\mathcal{U}_m$, $\mathcal{I}_n$ & the user set with the same identity $m$, the item set with the same product image $n$\\[0.5mm]
      market segment $(m,n)$ & the market defined for users with the same identity $m$ on products with the same type of image $n$ \\[0.5mm]
      $\mathcal{D}=\{r_{u,i}|u,i\}$ & the complete interaction data\\[0.5mm]
      $\mathcal{D}_{m,n}$, $\mathcal{D}_m$, $\mathcal{D}_n$ & interactions within the market segment $(m,n)$, $\mathcal{D}_m=\bigcup_n \mathcal{D}_{m,n}$, $\mathcal{D}_n=\bigcup_m \mathcal{D}_{m,n}$\\
    \bottomrule
    \end{tabular}
    \caption{Important terms and notation.}
    \label{tab:notation}
\end{table}

\xhdr{User Identity Group (Gender).}
Unfortunately, gender identities of Amazon users are not directly accessible.
We thus leverage users' interactions with \textit{Clothing} products in the \textit{Amazon 2018 Dataset} to access their gender identities, where most products are explicitly classified into \textit{Women's Clothing} or \textit{Men's Clothing}. As shown in the figure below, we find a clear bimodal distribution of purchase frequency towards gender-specific clothing products. 
\begin{center}
\includegraphics[width=0.6\linewidth]{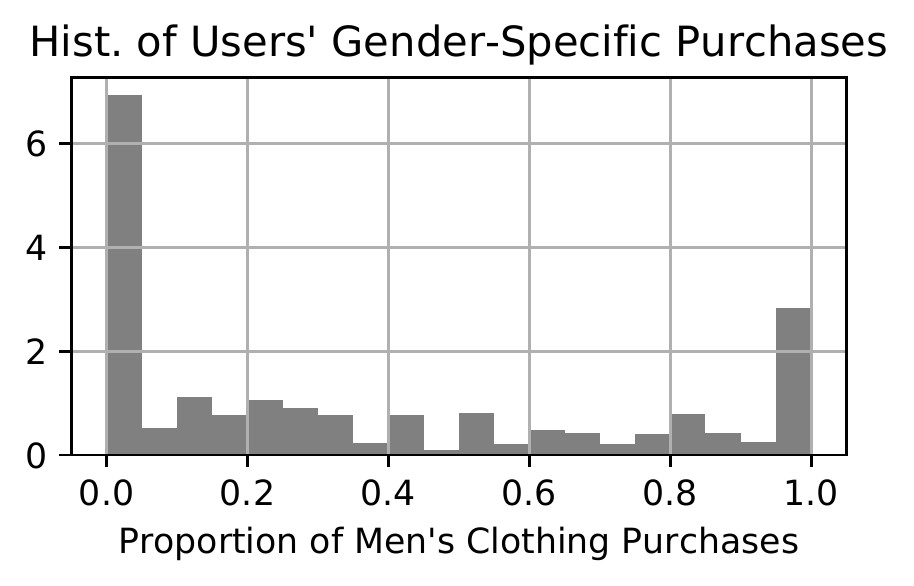}
\end{center}
We discard `ambiguous' users whose men's (or women's) clothing purchase frequencies fall into 40\%-60\%, and identify the remaining users as `Female' (69\%) or `Male' (31\%). Finally 11\% of total users in the \textit{Electronics} category can be matched to these identities and 53\% of them are identified as `Female'.

After removing products without any human models, we are still able to obtain a large-scale dataset containing around 1.3M rating scores across 9,560 electronics products from 1.1M users. Note that although the inferred product gender image and user identity are not as precise as in \dataname{ModCloth}, this dataset is dramatically different from \dataname{ModCloth} regarding its scale and sparsity. In contrast to the relationship between a user's interactions with clothing products and a dimension of \textit{human body shape}, we speculate that \textit{gender} is intuitively less relevant to the intrinsic qualities of most products in \dataname{Electronics}; thus its effects on users' interactions are possibly more likely to come from marketing bias.

\section{Statistical Analysis}\label{sec:analysis}
We split a consumer's product preference into two dimensions: 1) the user's preference in terms of willingness to consume (purchase) a product; and 2) the user's satisfaction feedback (e.g.~ratings) on the consuming experience.
We then conduct observational studies on the \dataname{ModCloth} and \dataname{Electronics} datasets to address marketing bias across the above two dimensions.
\begin{itemize}
    \item We first investigate if there is a bias introduced by a particular marketing effect in a consumer's \textit{product selection} process. Specifically, we examine if a correlation exists between product image and user identity in terms of interaction frequency in our datasets.
    \item Then we study \textit{consumer satisfaction} regarding the purchased products as a function of product image, user identity, and their second-order interactions. These consumer feedback signals include rating scores on \dataname{ModCloth} and \dataname{Electronics}, as well as the binarized fit feedback (i.e.,~if the clothing product fits the user) on \dataname{ModCloth}.
\end{itemize}
Important terms and notation throughout the paper are included in \cref{tab:notation}.

\subsection{Product Selection vs. Marketing Bias}
\begin{table}
\setlength{\tabcolsep}{4pt}
\begin{tabular}{lrrrrrr}
\toprule
& \multicolumn{3}{c}{ModCloth} & \multicolumn{3}{c}{Electronics}\\
\cmidrule(lr){2-4}
\cmidrule(lr){5-7}
{} &     $\chi^2$ &  p-value &      \#reviews &     $\chi^2$ &  p-value &      \#reviews\\
\cmidrule(lr){2-4}
\cmidrule(lr){5-7}
all    &  158.7 &    <0.001 &  91,526 &  581.8 &      <0.001 &  174,124 \\
\cmidrule(lr){2-4}
\cmidrule(lr){5-7}
<=2014 &    0.5 &    0.466 &  25,383 &  151.0 &      <0.001 &   49,699 \\
2015   &   66.7 &    <0.001 &  20,241 &  172.7 &      <0.001 &   46,891 \\
2016   &   70.8 &    <0.001 &  21,239 &   96.4 &      <0.001 &   43,907 \\
>=2017 &   29.0 &    <0.001 &  24,663 &  120.8 &      <0.001 &   33,627 \\
\bottomrule
\end{tabular}
\caption{Results from $\chi^2$ test of the two-way contingency tables on \dataname{ModCloth} and \dataname{Electronics}.}\label{tab:ch2test}
\end{table}

Because of the constraint of conducting real-world experiments with random assignments, we instead address marketing bias in  product selection by analyzing the association between product image and user identity in observed data with respect to interaction frequency. 
Our null hypothesis is that product image and user identity are statistically independent. Given this assumption, we expect to see lower deviations of their observed frequencies and the marginally expected values. Therefore the following Pearson's Chi-Squared Test Statistic can be used to test the association between these two variables in terms of frequency \citep{everitt1992analysis}:
\begin{equation}
    \chi^2=\sum_{m, n} \frac{(f_{m, n} - \mathbb{E} f_{m, n})^2}{\mathbb{E}f_{m, n}},
\end{equation}
where $m$ and $n$ represent a user identity group and a product image group respectively, $f_{m,n}$ is the observed number of interactions in the market segment $(m,n)$ and $\mathbb{E} f_{m, n} = \frac{(\sum_{m'} f_{m', n})(\sum_{n'} f_{m, n'})}{(\sum_{m',n'}f_{m', n'})}$ represents its expectation. The null hypothesis will be rejected (i.e.,~the association between two variables exists in terms of frequency) if an extremely large $\chi^2$ is obtained (i.e.,~small $p$-value).

\begin{table}
\centering
\begin{subtable}{\linewidth}
\centering
\begin{tabular}{lllr}
\toprule
 & \multicolumn{2}{c}{\textbf{User Identity}} & \\
\textbf{Product Image} &  Small &  Large &    All \\
\cmidrule(lr){2-3}
Small       &  \cellcolor{lightlightgray}31,800 (+754.98) &   7,038 (-754.98) &  38,838 \\
Small\&Large &  41,361 (-754.98) &  \cellcolor{lightlightgray}11,327 (+754.98) &  52,688 \\
\cmidrule(lr){2-3}
All         &  73,161 &  18,365 &  91,526 \\
\bottomrule
\end{tabular}
\caption{ModCloth}\label{tab:contigency_modcloth}
\end{subtable}
\begin{subtable}{\linewidth}
\centering
\setlength{\tabcolsep}{2.5pt}
\begin{tabular}{lllr}
\toprule
 & \multicolumn{2}{c}{\textbf{User Identity}} & \\
\textbf{Product Image} &  Female &   Male &     All \\
\cmidrule(lr){2-3}
Female      &   \cellcolor{lightlightgray}34,259 (+1,472.89) &  31,587 (-1,472.89) &   65,846 \\
Female\&Male &   26,478 (+880.88) &  24,930 (-880.88) &   51,408 \\
Male        &   25,963 (-2,353.77) &  \cellcolor{lightlightgray}30,907 (+2,353.77) &   56,870 \\
\cmidrule(lr){2-3}
All         &   86,700 &  87,424 &  174,124 \\
\bottomrule
\end{tabular}
\caption{Electronics}\label{tab:contigency_electronics}
\end{subtable}
\caption{Contingency tables of the frequency distribution of product images and user identities on \dataname{ModCloth} and \dataname{Electronics}. Deviations ($f_{m, n} - \mathbb{E} f_{m, n}$) from the expected frequency values are provided in parentheses.}\label{tab:contingency}
\end{table}

To further separate the potential marketing bias from trending effects, we conduct association tests on the complete interaction data as well as interactions within different time spans. Test results are included in \cref{tab:ch2test}, where we find all $p$-values are smaller than 0.001 except for the test on interaction data before 2014 on \dataname{ModCloth}. 
These results may imply the existence of the association between product image and user identity in consumers' product selections. 

In \cref{tab:contingency}, we provide contingency tables of the frequency distribution of different market segments and their deviations from expected values ($f_{m,n}-\mathbb{E} f_{m, n}$). We observe generally more interactions than 
expected
on the consumer-product segments where users' identities match the product images (`self-congruity'), while several market segments are underrepresented in the data. For example, (`Large' user, `Small' product) on \dataname{ModCloth} and (`Female' user, `Male' product) on \dataname{Electronics} have smaller market sizes compared with other market segments.

\subsection{Consumer Satisfaction vs. Marketing Bias}\label{sec:ftest}
\begin{table}
\setlength{\tabcolsep}{3.5pt}
\begin{tabular}{lrrrrrr}
\toprule
& \multicolumn{4}{c}{ModCloth} & \multicolumn{2}{c}{Electronics} \\
\cmidrule(lr){2-5}
\cmidrule(lr){6-7}
& \multicolumn{2}{c}{Rating} & \multicolumn{2}{c}{Fit} & \multicolumn{2}{c}{Rating}\\
{} &   F-stat &  p-value &   F-stat &  p-value &   F-stat &  p-value \\
\cmidrule(lr){2-3}
\cmidrule(lr){4-5}
\cmidrule(lr){6-7}
product      &  171.9 &      <0.001 &  293.1 &    <0.001&  62.6 &    <0.001 \\
user         &   46.3 &      <0.001 &  402.4 &    <0.001&    3.5 &    0.061 \\
user$\times$product &   30.7 &      <0.001 &    0.0 &    0.997&    0.9 &    0.404 \\
\bottomrule
\end{tabular}
\caption{Results from two-way analysis of variance (ANOVA) on \dataname{ModCloth} and \dataname{Electronics}.}\label{tab:ftest}
\end{table}

Next we investigate consumer satisfaction as a function of product image and user identity through a standard statistical technique: two-way analysis of variance (\textbf{ANOVA}) \citep{kleinbaum1988applied}. 
We use rating scores to represent users' satisfaction regarding the overall quality of their consuming experience on both \dataname{ModCloth} and \dataname{Electronics}. For \dataname{ModCloth}, we also study consumer satisfactions with respect to their fit feedback (where `Just Right' is regarded as positive while all others are regarded as negative). The two-way \textbf{ANOVA} model can be formulated as
\begin{equation*}
    \mathit{consumer}\ \mathit{satisfaction} \sim \mathit{product} + \mathit{user}
    + \mathit{product}\times \mathit{user},
\end{equation*}
where the null hypotheses of our tests include
\begin{itemize}[leftmargin=13pt]
    \item[(a)] the average consumer satisfaction is equal across different product image groups;
    \item[(b)] the average consumer satisfaction is equal across different consumer identity groups;
    \item[(c)] there is no interaction effect between product groups and consumer groups with respect to satisfaction.
\end{itemize}
Given these assumptions, we may expect a lower variance of average satisfactions across different groups (\textit{between-group variation}) compared with the summation of satisfaction variations within each group (\textit{within-group variation}). Therefore, the standard \textbf{F-statistic}, defined as the between-group variation divided by the within-group variation \citep{kleinbaum1988applied}, can be applied to evaluate the correlations.

\begin{figure}
	\centering
	\begin{subfigure}[b]{0.295\linewidth}
		\centering
		\includegraphics[width=\textwidth]{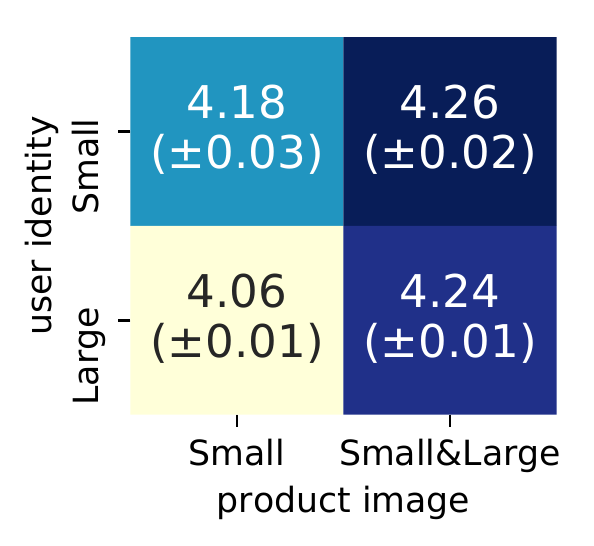}
		\caption{ModCloth (Rt.)}\label{fig:modcloth_rating_heatmap}
	\end{subfigure}%
	~ 
	\begin{subfigure}[b]{0.295\linewidth}
		\centering
		\includegraphics[width=\textwidth]{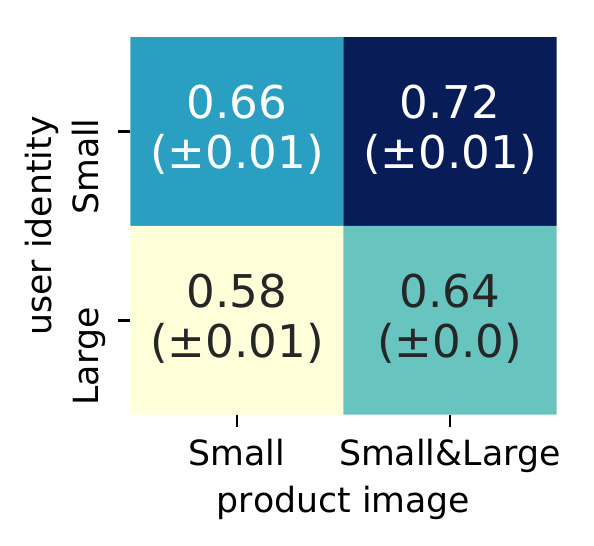}
		\caption{ModCloth (Fit)}\label{fig:modcloth_fit_score_heatmap}
	\end{subfigure}
	~ 
	\begin{subfigure}[b]{0.4\linewidth}
		\centering
		\includegraphics[width=\textwidth]{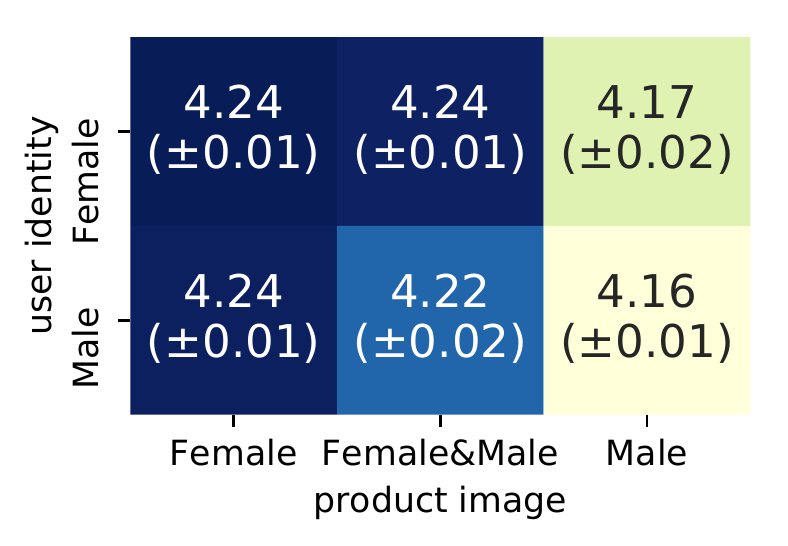}
		\caption{Electronics (Rating)}\label{fig:electronics_rating_heatmap}
	\end{subfigure}
	\caption{Heatmaps of sample means within market segments regarding (a) rating scores on \dataname{ModCloth}, (b) fit feedback on \dataname{ModCloth} and (c) rating scores on \dataname{Electronics}.}\label{fig:heatmaps}
\end{figure}

Results from statistical tests are included in \cref{tab:ftest}. The heatmaps of sample means within market segments and their 95\% confidence intervals are provided in \cref{fig:heatmaps}. 
We observe that users' rating scores are significantly different across market segments on \dataname{ModCloth}. For example `Large' users provide lower ratings on `Small' products (\cref{fig:modcloth_rating_heatmap}). Although users' fit feedback differs across product groups and user groups (hypothesis (a) and (b) are rejected in \cref{tab:ftest}), 
their association regarding fit feedback is negligible 
(results for `user$\times$product' in \cref{tab:ftest}). 
According to \cref{fig:modcloth_fit_score_heatmap}, we find clothing products in the \dataname{ModCloth} dataset generally fit better on `Small' users, and those products represented by human models with different body shapes (`Small\&Large') tend to obtain better fit feedback.
Although the `self-congruity' pattern is significant in the product selection process on \dataname{Electronics} (see \cref{tab:ch2test}, \cref{tab:contigency_electronics}), 
the interaction between product `gender' and user gender is insignificant with respect to users' rating scores (user$\times$product in \cref{tab:ftest}).

\subsection{Summary of Observations}
We summarize insights obtained from the above statistical analysis as follows:
\begin{itemize}
    \item The association between product image and user identity is consistently significant in terms of frequency distribution, implying
    the existence of marketing bias in the collected interaction datasets. The `self-congruity' pattern is also observable, i.e., consumers may generally tend to interact with products with similar impressions as their identities. Such an association notably causes underrepresentation of certain market segments.
    \item The relationship between consumer satisfaction and marketing factors is rather complicated. We observe rating disparities across
    product groups and user groups,
    while the existence of their interaction effect depends on the type of product and the type of satisfaction measure. We find a similar `self-congruity' pattern for rating scores on \dataname{ModCloth} while the `user$\times$product' term remains insignificant in the other two testing scenarios. 
\end{itemize}

\section{Market-Fairness of Recommender Systems}\label{sec:fair}
From the above analysis, we have
confirmed that our interaction data is correlated to (and possibly affected by) marketing strategies used by product retailers (i.e.,~selections of human models). Our next step is to study if (and how) this marketing bias is propagated by algorithms from input data to recommendation results.

\xhdr{Problem Setting.} In this study, we focus on recommendation algorithms trained on explicit feedback (i.e.,~rating scores). The primary predictive task is formulated as a \textit{rating prediction problem}: rating scores ($r_{u,i}$) are assumed to reflect users' preferences over products, and algorithms are trained to generate users' product preference scores ($s_{u,i}$) which approximate these ratings.

Unlike previous studies \citep{beutel2019fairness,DBLP:conf/aies/2019,zhu2018fairness,yao2017beyond} which focus on evaluating and protecting the fairness of a single side (user or product) of recommender systems, in the context of marketing bias, we are particularly interested in the global market fairness of the recommendations, i.e.,~user-fairness and product-fairness need to be protected at the same time. Specifically we describe the market fairness in the explicit feedback setting along 
two dimensions.
\begin{itemize}[leftmargin=13pt]
    \item[(a)] Averaged errors of rating predictions from a recommendation algorithm across different consumer-product market segments are expected to be equal. 
    \item[(b)] The distribution of market segments in terms of frequency within recommended interactions are expected to be consistent with the distribution within the real interaction data.
\end{itemize}

\xihdr{Rating Prediction Fairness.} 
We notice that the first market fairness description is indeed consistent with the null hypothesis of a one-way \textbf{ANOVA} test about the association between prediction errors ($e_{u,i}=s_{u,i} - r_{u,i}$) and market segments ($(m,n)$). That is, with the assumption that average prediction errors from a fair algorithm are supposed to be irrelevant to market segments, we expect to observe a lower variation of average errors across market segments (\textit{bewteen-segment variation}) compared to the error variations within each segment (\textit{within-segment variation}). Specifically these variations can be defined as
\begin{equation*}
\begin{aligned}
\text{between-segment var.:}\quad &V^{\mathit{(market)}}=\frac{1}{|\mathcal{D}|}\sum_{m,n} |\mathcal{D}_{m,n}|\left(\bar{e}_{m,n,\cdot} - \bar{e}\right)^2\\
\text{within-segment var.:}\quad &U^{\mathit{(market)}}=\frac{1}{|\mathcal{D}|}\sum\limits_{m,n}~\sum\limits_{\substack{u\in\mathcal{U}_m,\\ i\in\mathcal{I}_n}} \!\!\!\!\! \left(e_{u,i} - \bar{e}_{m,n,\cdot}\right)^2\\
\end{aligned}
\end{equation*}
where $\bar{e}_{m,n,\cdot}$ denotes the sample mean of prediction errors within the market segment $(m,n)$; $|\mathcal{D}_{m,n}|$ represents the number of interactions included in a consumer-product segment $(m,n)$; $|\mathcal{D}|$ denotes the total sample size. 

To ensure a tractable distribution for significance testing, the above two terms are corrected by their degrees-of-freedom and the following \textbf{F-statistic} can thus be calculated:
\begin{equation}
    \mathcal{F}^{\mathit{(market)}} = \frac{V^{\mathit{(market)}}~/~(M\times N - 1)}{U^{\mathit{(market)}}~/~\underbrace{(|\mathcal{D}| - M\times N)}_{\mathclap{\text{correction of degree of freedom}}}} \label{eq:feval}.
\end{equation}
Then we obtain a fairness evaluation metric to evaluate a global parity of prediction errors across different consumer-product market segments, where lower $\mathcal{F}$ indicates better rating prediction fairness. 

\xihdr{Product Ranking Fairness.} We further investigate the fairness of the product ranking performance from recommendation algorithms. 
For each user, we rank all products based on the predicted preference scores $s_{u,i}$ and regard the top-ranked $K$ items as recommended products. 
By gathering users and the recommended products,
we are able to obtain the frequency distribution of market segments within these predicted interactions $p_{m,n}\sim\mathcal{P}$. We regard the frequency distribution of market segments in the real interactions $q_{m,n}\sim\mathcal{Q}$ as the reference distribution, and evaluate the deviation of $\mathcal{P}$ from $\mathcal{Q}$ using the following \textbf{KL-divergence} \citep{kullback1951information}:
\begin{equation}
    D_{\mathit{KL}}(\mathcal{P} | \mathcal{Q}) = \sum_{m,n} p_{m,n}\log\left(\frac{p_{m,n}}{q_{m,n}}\right).\label{eq:kl}
\end{equation}
We use this metric to evaluate the product ranking fairness. Lower $D_{\mathit{KL}}$ indicates better fairness.

\section{A Fairness-Aware Framework}\label{sec:fairMF}
A common optimization criterion for model-based collaborative filtering algorithms in the explicit feedback setting is based on \textbf{MSE}, i.e.,~minimizing the following loss function
\begin{equation}
    \mathcal{L} = \sum (s_{u,i} - r_{u,i})^2.
\end{equation}
A popular choice to model the preference score $s_{u,i}$ is through matrix factorization \citep{koren2009matrix}
\begin{equation}
    s_{u,i} = b_0 + b_i + b_u + \left<\bm{\gamma}_i,  \bm{\gamma}_u\right>.\label{eq:mf}
\end{equation}
In \cref{eq:mf}, $b_0$ is the global intercept, $b_i$ and $b_u$ are item-specific and user-specific offsets, $\bm{\gamma}_i$ and $\bm{\gamma}_u$ are $d$-dimensional embeddings to capture items' latent properties and users' latent preferences on these dimensions. 

\xhdr{Error Correlation Loss.} 
Following previous work using the regularizing schemes \citep{yao2017beyond,beutel2019fairness,abdollahpouri2017controlling},
we propose a fairness-aware framework by considering an error correlation loss to regularize systematic error biases on the market:
\begin{equation}
    \mathcal{L}^* = \sum(s_{u,i} - r_{u,i})^2 + \alpha \mathcal{L}_{\mathit{corr.}},
\end{equation}
where $\mathcal{L}_{\mathit{corr.}}$ is an additional term to regularize the correlation between prediction errors $e_{u,i}$ and the distribution of market segments $(m,n)$. $\alpha$ is a hyperparameter to control the trade-off between prediction accuracy and this correlation penalty term. 

In practice, we consider the following form by relaxing the evaluation metric \cref{eq:feval}:
\begin{equation}
    \mathcal{L}_{\mathit{corr.}} = \kappa^{(\mathit{u.})} \overbrace{\frac{V^{(\mathit{u.})}}{U^{(\mathit{u.})}}}^{\mathclap{\text{error parity on user identity}}} +  \kappa^{(\mathit{p.})} \underbrace{\frac{V^{(\mathit{p.})}}{U^{(\mathit{p.})}}}_{\mathclap{\text{error parity on product image}}} +  \kappa^{\mathit{(market)}} \overbrace{\frac{V^{\mathit{(market)}}}{U^{\mathit{(market)}}}}^{\mathclap{\text{error parity on market segments}}}, \label{eq:corr}
\end{equation}
where $V^{(\mathit{u.})}, U^{(\mathit{u.})}$, $V^{(\mathit{p.})}, U^{(\mathit{p.})}$ can be implemented by merging market segments within the same type of user identity groups or product image groups.
Note that the three error parity terms in \cref{eq:corr} can be regarded as simplified implementations of the fairness metric in \cref{eq:feval}.
$\kappa^{(\mathit{u.})}, \kappa^{(\mathit{p.})}, \kappa^{\mathit{(market)}} \in \{0,1\}^3$ are binary hyperparameters to instantiate different forms of correlation loss. For example, a selection of $(\kappa^{(\mathit{u.})}, \kappa^{(\mathit{p.})}, \kappa^{\mathit{(market)}}) = (1, 0, 0)$ represents that we only penalize the correlation between prediction errors and user identity groups.

\section{Experiments}
We conduct experiments on the collected \dataname{ModCloth} and \dataname{Amazon} datasets to evaluate the recommendation performance and the market fairness as described in \cref{sec:fair}.

\xhdr{Baselines.} The following standard algorithms are considered:
\begin{itemize}
    \item \textbf{itemCF}, an item-based collaborative filtering algorithm \citep{sarwar2001item,linden2003amazon};
    \item \textbf{userCF}, a user-centric collaborative filtering method \citep{herlocker1999algorithmic};
    \item \textbf{MF}, the matrix factorization method \citep{koren2009matrix}, where the value of the preference prediction $s_{u,i}$ is unbounded;
    \item \textbf{PoissonMF}, a hierarchical Bayesian framework where the preference factorization is linked to the rating score through a Poisson distribution, so that the preference score $s_{u,i}$ is bounded as a positive value \citep{gopalan2015scalable}.
\end{itemize}
By studying the recommendation outputs from these methods, we evaluate how standard collaborative filtering algorithms respond to the marketing bias in the input data.

We implement our proposed framework (\textbf{MF (corr.error)}), where $s_{u,i}$ is factorized using matrix factorization. By comparing its performance with the above methods (especially \textbf{MF}), we evaluate if the rating prediction and the product ranking fairness can be improved without losing much accuracy by adding the proposed correlation loss. 
Besides, we consider another two fairness-aware alternatives:
\begin{itemize}
    \item \textbf{MF (corr.value)}, a method similar to \textbf{MF (corr.error)} except that $\mathcal{L}_{\mathit{corr.}}$ is implemented as the correlation between the predicted rating \textit{values} $s_{u,i}$  and the market segments. By comparing \textbf{MF (corr.error)} with it, we evaluate the effectiveness of controlling the parity of prediction errors instead of the absolute statistical parity of prediction values.
    
    \item \textbf{MF (reweighted)}, a method where the loss function is reweighted based on the sizes of market segments in the training data. We also consider the following generic form of the loss function:
    \begin{equation*}
        \mathcal{L} =  \frac{\kappa^{(\mathit{u.})}}{M}\sum_m \mathit{MSE}_m +   \frac{\kappa^{(\mathit{p.})}}{N}\sum_n \mathit{MSE}_n +  \frac{\kappa^{\mathit{(market)}}}{M N}\sum_{m,n} \mathit{MSE}_{m,n}.
    \end{equation*}
     By comparing it with other baselines, we study if the marketing bias can be alleviated by simply increasing the weights of underrepresented segments in the training data.
\end{itemize}
For all above methods, we primarily evaluate their rating prediction accuracy through \textbf{MSE} and \textbf{MAE}, and rating prediction fairness in terms of the \textbf{F-statistic} (\cref{eq:feval}). We also evaluate their recommendation accuracy through \textbf{AUC} and \textbf{NDCG}, and the product ranking fairness in terms of \textbf{KL-divergence} (\cref{eq:kl}).

\xhdr{Experimental Details.} We use the following rules to split interactions into train/validation/test sets: for users with at least two reviews, their most recent ratings are regarded as a test set; for users with at least three reviews, their second-to-last ratings are used for validation; the remaining interactions are used for training. We apply the same analysis on both training and test sets as in \cref{sec:analysis}, and find similar patterns except that fewer female users (40\%) are included in the test set of \dataname{Electronics}.\footnote{85\% female users (vs.~73\% male users) have only one review in our entire dataset.} 

We use the \textbf{ADAM} optimizer \citep{DBLP:conf/iclr/2015} with a learning rate of 0.001, a batch size of 512 and a fixed dimensionality of the latent embeddings in all model-based methods ($d=10$). An $\ell_2$ regularizor is applied on all model-based methods, where $\lambda$ is selected from $\{0.01, 0.1, 1.0, 10\}$. The accuracy-fairness trade-off $\alpha$ is chosen from $\{0.5, 1.0, 5.0, 10.0\}$. All hyperparameters are selected based on the recommendation accuracy\footnote{\textbf{MSE} for rating prediction accurcy and \textbf{NDCG} for product ranking.} on the validation set. 
For fairness-aware methods, we search hyperparameters $\bm{\kappa}=(\kappa^{(\mathit{u.})}, \kappa^{(\mathit{p.})}, \kappa^{\mathit{(market)}})$ from $\{(1,0,0), (0,1,0), (0,0,1), (1,1,0), (1,1,1)\}$. For each $\bm{\kappa}$, we first decide all other hyperparameters based on their recommendation accuracy, then select $\bm{\kappa}$ which yields the fairest recommendation results on the validation set. For each user, the top-10 ranked products are regarded as recommended items. Reviews in the test set where rating scores are larger than 3 are considered as reference interactions for the ranking task. All results are reported on the test set.

\subsection{How does a standard collaborative filtering algorithm respond to biased input data?}

We report the above mentioned rating prediction and product ranking metrics on \dataname{ModCloth} and \dataname{Electronics}, regarding both accuracy and fairness, in \cref{tab:results}. We first investigate standard recommendation methods without any explicit fairness controls (i.e.,~\textbf{itemCF}, \textbf{userCF}, \textbf{PoissonMF} and \textbf{MF}). We observe that most methods yield biased prediction results on both datasets according to the \textbf{F-statistic}-based significance test. Although we find seemingly fair prediction errors from \textbf{userCF}, it actually produces a much larger \textbf{MSE} (as well as worse product ranking results) compared to other methods. 

\begin{figure}
	\centering
	\begin{subfigure}[b]{0.445\linewidth}
		\centering
		\includegraphics[width=\textwidth]{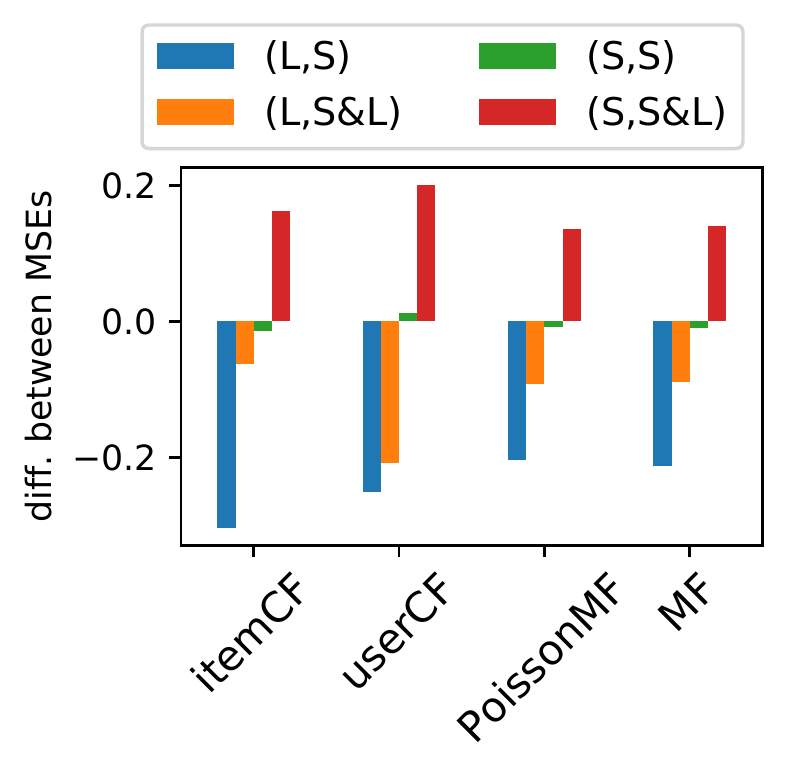}\vspace{-0.1in}
		\caption{ModCloth}\label{fig:pairwise_modcloth}
	\end{subfigure}%
	~ 
	\begin{subfigure}[b]{0.555\linewidth}
		\centering
		\includegraphics[width=\textwidth]{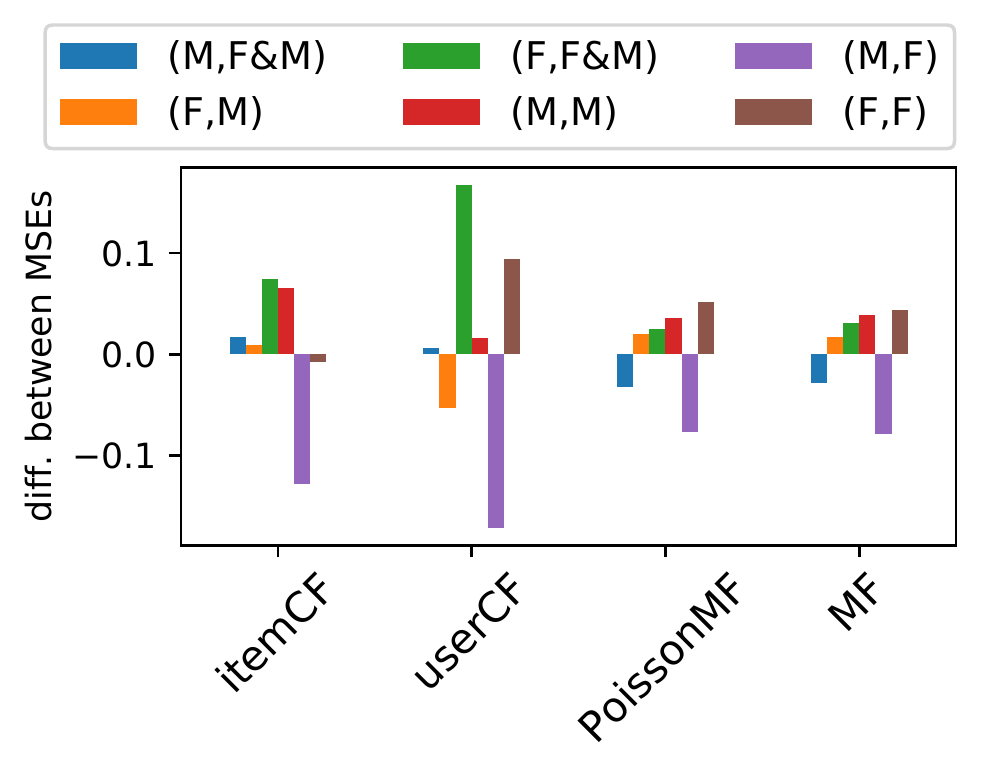}\vspace{-0.1in}
		\caption{Electronics}\label{fig:pariwise_electronics}
	\end{subfigure}
	\caption{Differences between the out-segment \textbf{MSE}s and the in-segment \textbf{MSE}s. Market segments are sorted based on their market sizes in the training data.
	}\label{fig:pairwise_fairness}
\end{figure}

We further calculate the differences between the out-segment \textbf{MSE}s and the in-segment \textbf{MSE}s for these algorithms. Given a market segment $(m,n)$, we have
\begin{equation}
    \mathit{diff}_{m,n} = \mathit{MSE}_{u\notin\mathcal{U}_m \mathit{or} i\notin\mathcal{I}_n} - \mathit{MSE}_{u\in\mathcal{U}_m \mathit{and} i\in\mathcal{I}_n}.
\end{equation}
$\mathit{diff}_{m,n}>0$ indicates, for an algorithm, the market segment $(m,n)$ is more predictable (smaller \textbf{MSE}) than the interactions outside it. These differences are displayed in \cref{fig:pairwise_fairness}, where the market segments are sorted based on their training sizes. We observe an overall trend that all algorithms generally tend to favor the dominating market segments (e.g.~`Small' users on `Small\&Large' products in \dataname{ModCloth}) in varying degrees. 
We find the correlation between the predictibility and the market segment size is more prominent on \dataname{ModCloth} but rather complicated on \dataname{Electronics}. 
However, by cross matching \cref{fig:pairwise_fairness} and the contingency table \cref{tab:contigency_electronics}, we find that the trend correlates to the deviations of the real market size and the expected market size: the consumer-product segments (`Female', `Male'), (`Male', `Female') and (`Male', `Female\&Male') are underrepresented based on this difference ($f_{m,n} - \mathbb{E}f_{m,n}<0$), also generally unfavored by the recommendation algorithms.

\begin{figure}
	\centering
	\begin{subfigure}[b]{0.395\linewidth}
		\centering
		\includegraphics[width=\textwidth]{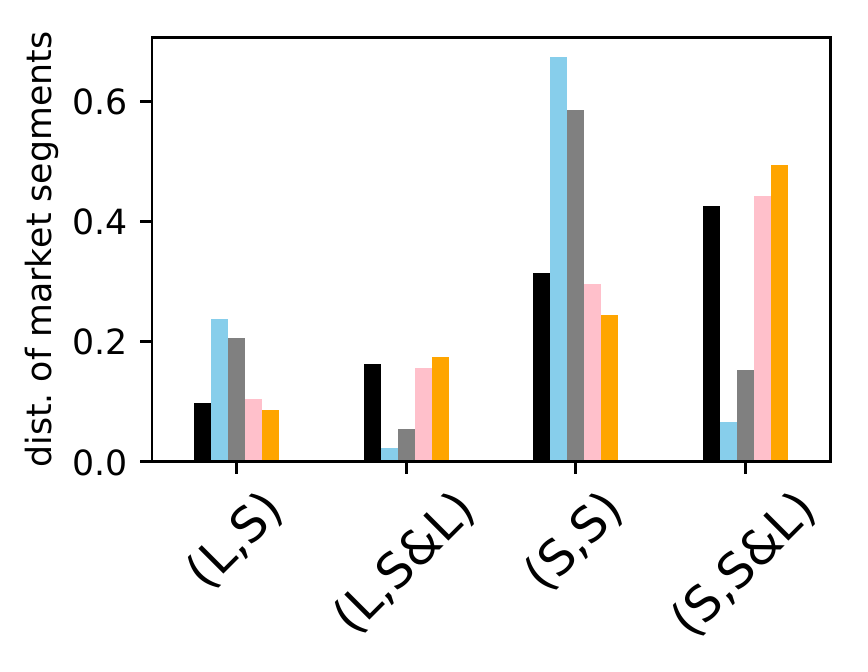}
		\caption{ModCloth}\label{fig:ranking_modcloth}
	\end{subfigure}%
	~ 
	\begin{subfigure}[b]{0.605\linewidth}
		\centering
		\includegraphics[width=\textwidth]{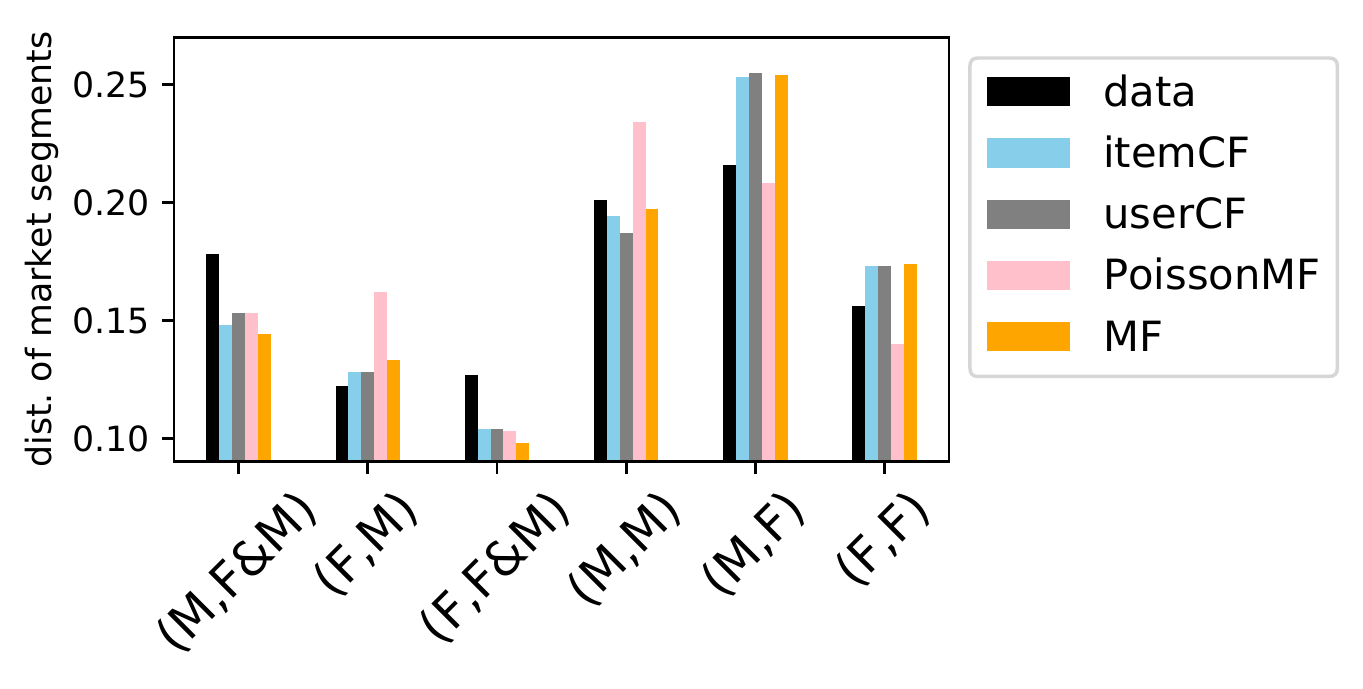}
		\caption{Electronics}\label{fig:ranking_electronics}
	\end{subfigure}
	\caption{Distribution of market segments within test data (positive interactions only) and within recommendations. Market segments are sorted based on their sizes in training data.
	}\label{fig:rank_fairness}
\end{figure}

We display the distributions of market segments within positive interactions where rating scores are larger than 3 and the recommended top-10 products from these algorithms in \cref{fig:rank_fairness}. Compared with the distributions in real interactions (the `data' columns in \cref{fig:rank_fairness}), we can observe the deviations of recommendation results from most algorithms, particularly \textbf{itemCF} and \textbf{userCF} on \dataname{ModCloth}. However, systematic patterns about how these deviations correlate to the sizes of different market segments in the training data are not observed.

\subsection{Can recommendation fairness be improved by applying the correlation loss?}
\begin{figure}
	\centering
	\begin{subfigure}[b]{0.52\linewidth}
		\centering
		\includegraphics[width=\textwidth]{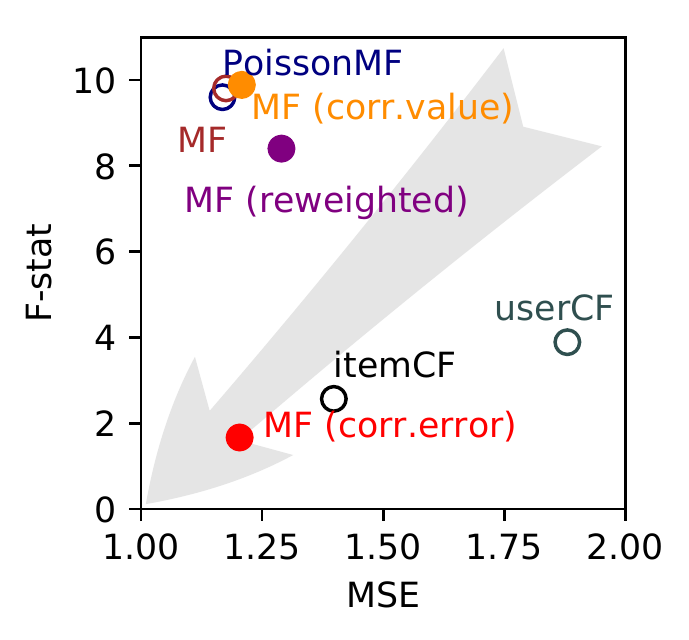}
		\caption{ModCloth (Rating)}\label{fig:}\vspace{0.1in}
	\end{subfigure}%
	~ 
	\begin{subfigure}[b]{0.48\linewidth}
		\centering
		\includegraphics[width=\textwidth]{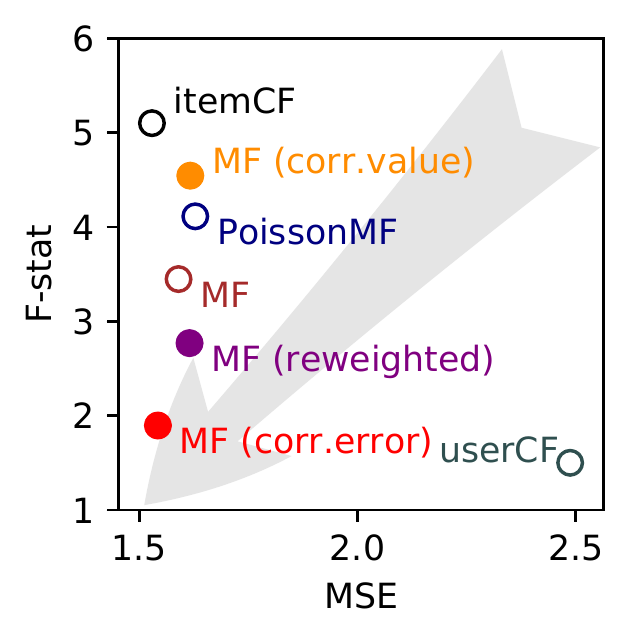}
		\caption{Electronics (Rating)}\label{fig:}\vspace{0.1in}
	\end{subfigure}\\
	\begin{subfigure}[b]{0.49\linewidth}
		\centering
		\includegraphics[width=\textwidth]{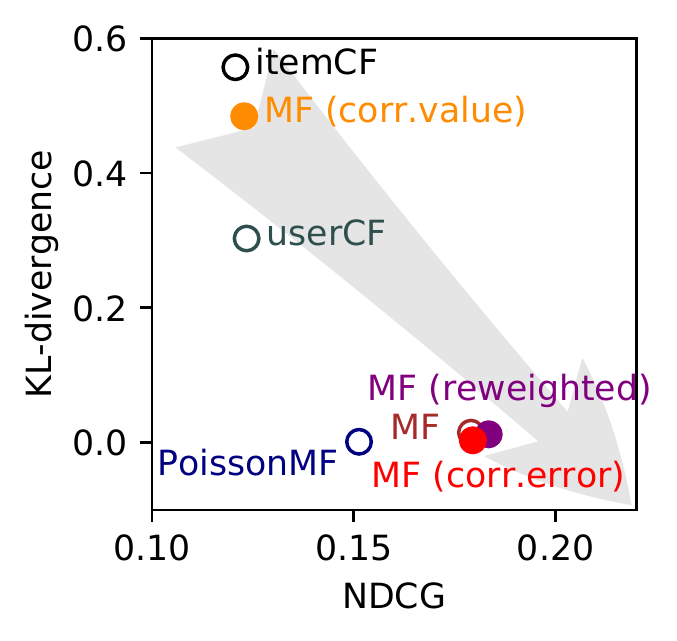}
		\caption{ModCloth (Ranking)}\label{fig:}\vspace{0.1in}
	\end{subfigure}%
	~ 
	\begin{subfigure}[b]{0.51\linewidth}
		\centering
		\includegraphics[width=\textwidth]{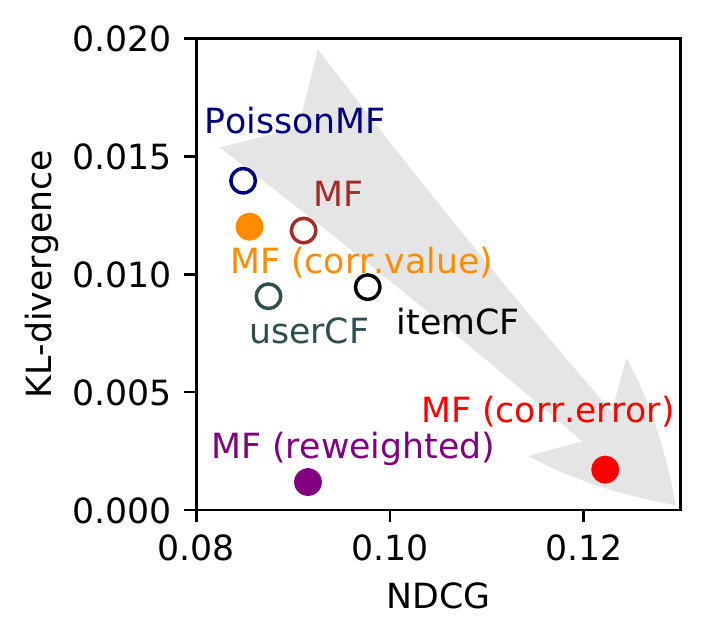}
		\caption{Electronics (Ranking)}\label{fig:}\vspace{0.1in}
	\end{subfigure}
	\caption{Scatter plots for accuracy-fairness trade-off from different algorithms. 
	Shaded arrows indicate the most ideal direction: higher accuracy, better fairness.}\label{fig:tradeoff}
\end{figure}

In \cref{tab:results}, we further compare the results from fairness-aware algorithms in group (b) to the standard algorithms in group (a), particularly \textbf{MF}. To better visualize the trade-off between recommendation accuracy and market fairness, we present scatter plots of an accuracy metric and a fairness metric on both datasets in \cref{fig:tradeoff}. We notice the proposed method with error correlation loss \textbf{MF (corr.error)} generally provides better rating and ranking fairness (lower \textbf{F-statistic} and \textbf{KL-divergence}) than standard \textbf{MF}, without trading-off much recommendation accuracy. An interesting finding is the combination selection $\bm{\kappa}$ on the validation set is consistent with our analysis in \cref{tab:ftest}: the complete correlation loss ($\bm{\kappa}=(1,1,1)$) is selected for \dataname{ModCloth} and the addition of product and user correlation ($\bm{\kappa}=(1,1,0)$) is selected for \dataname{Electronics}.

\begin{table*}
    \centering
    \setlength{\tabcolsep}{4.5pt}
\begin{tabular}{clrrrrrrrrrrrrrr}
\toprule
& & \multicolumn{7}{c}{ModCloth} & \multicolumn{7}{c}{Electronics} \\
\cmidrule(lr){3-9}
\cmidrule(lr){10-16}
&  & \multicolumn{4}{c}{Rating Prediction} & \multicolumn{3}{c}{Product Ranking} & \multicolumn{4}{c}{Rating Prediction} & \multicolumn{3}{c}{Product Ranking} \\
& Method &    MSE &    MAE &  F-stat &  p-value & AUC &   NDCG &      KL &    MSE &    MAE &  F-stat &  p-value & AUC &   NDCG &      KL \\
\cmidrule(lr){1-2}
\cmidrule(lr){3-6}
\cmidrule(lr){7-9}
\cmidrule(lr){10-13}
\cmidrule(lr){14-16}
\multirow{4}{*}{(a)} & itemCF     &  1.398 &  \underline{0.841} &    2.568 &    0.053 &  0.601 &   0.121 &  0.557 &  \underline{1.529} &  \underline{0.966} &    5.099 &    <0.001 &  0.619 &  0.098 &  0.009 \\
& userCF     &  1.880 &  0.946 &    3.889 &    0.009 &  0.504 &  0.123 &  0.303 &  2.487 &  0.980 &    \underline{1.501} &    0.186 &  0.503 &  0.087 &  0.009 \\
& PoissonMF         &  \underline{1.168} &  0.859 &    9.600 &    <0.001 &  0.638 &  0.151 &  \underline{0.001} &  1.628 &  1.035 &    4.112 &    0.001 &  0.565 &  0.085 &  0.014 \\
\rowcolor{lllgray}
& MF          &  1.176 &  0.859 &    9.805 &    <0.001 &  0.817 &  0.179 &  0.015 &  1.590 &  1.025 &    3.447 &    0.004 &  0.591 &  0.091 &  0.012 \\
\cmidrule(lr){3-6}
\cmidrule(lr){7-9}
\cmidrule(lr){10-13}
\cmidrule(lr){14-16}
\multirow{3}{*}{(b)} & MF (reweighted) &  1.290 &  0.872 &    8.402 &    <0.001 &  \underline{0.852} &  \underline{0.183} &  0.012 &  1.615 &  1.017 &    2.769 &    0.017 &  0.594 &  0.092 &  \underline{0.001} \\
& MF (corr.value)       &  1.208 &  0.875 &    9.887 &    <0.001 &  0.549 &  0.123 &  0.484 &  1.617 &  1.043 &    4.543 &    <0.001 &  0.502 &  0.086 &  0.012 \\
\rowcolor{lllgray}
& MF (corr.error)       &  1.204 &  0.873 &    \underline{1.667} &    0.172 &  0.818 &  0.179 &  0.003 &  1.543 &  1.011 &    1.896 &    0.091  &  \underline{0.766} &  \underline{0.122} &  0.002 \\
\bottomrule
\end{tabular}
    \caption{Recommendation results on \dataname{ModCloth} and \dataname{Electronics}. 
    For rating predictions, \textbf{MSE},\textbf{MAE} are used to evaluate the prediction accuracy while the \textbf{F-statsitic} (\cref{eq:feval}) is used to evaluate the prediction fairness and its associated \textbf{p-value} is provided; for product rankings, \textbf{AUC} and \textbf{NDCG} are used to evaluate the recommendation accuracy while the \textbf{KL-divergence} (\cref{eq:kl}) is used to evaluate the recommendation fairness. 
    The most accurate and the fairest results are \underline{underlined}.}
    \label{tab:results}
\end{table*}

We find the reweighting scheme also benefits the fairness metrics, particularly in the product ranking setting.
One surprising finding is that by applying the error correlation loss, a significant performance gain in terms of product ranking accuracy (\textbf{AUC} and \textbf{NDCG}) can be obtained on \dataname{Electronics}. A possible reason could be that \dataname{Electronics} is an extremely sparse dataset where algorithms like \textbf{MF} may struggle to converge to an ideal local optimum. The fairness-aware correlation loss, however, could help regularize the training process.

\section{Conclusions and Future Work}
We conclude our work and summarize 
our findings
as follows:
\begin{itemize}
    \item We investigated a potential source of bias---marketing bias---in the form of the association between interaction feedback, product image and user identity, on two real-world e-commerce datasets. Through observational studies, the inter-correlations between these factors can be confirmed and the `self-congruity' patterns are noticeable in the product selection process, which eventually results in the underrepresentation of some market segments.
    \item We focused on market fairness and investigated how standard collaborative filtering algorithms react to this biased input data. We found such a bias can be propagated to the recommendation outcomes in varying degrees.
    \item We developed an error correlation framework, which explicitly calibrates the equity of prediction errors across different market segments. Experimental results demonstrate that by applying this correlation loss, a superior accuracy-fairness trade-off can be achieved.
\end{itemize}
This work is a first step to approach the potential marketing bias in machine learning systems. We also wish to address several limitations of our data and methods, and to provide potential research directions. 
\begin{itemize}
\item \xihdr{Data.} We study marketing bias by formulating it as the relationship between the human model images of products and user identities. Multiple marketing factors (e.g.~product descriptions, social media advertisement contents) can also be considered. Binary gender identities are inferred in our \dataname{Electronics} dataset, which is limited to represent user identities that are not exclusively masculine or feminine, e.g.~users who don't always purchase products corresponding to their own identities, or those who identify themselves outside the binary definition.
\item \xihdr{Analysis.} We collect \dataname{ModCloth} and \dataname{Electronics} as logged interactions where many confounders (inventory status, the observability of each product, potential biases introduced in the scraping and preprocessing stage, etc.) exist and are difficult to be disentangled. Although the inter-correlation between product image and user identity is observed in these datasets, we cannot draw any causal conclusions without controlling some notable confounding factors. Therefore another direction to validate (or more fundamentally address) this marketing bias is to conduct user-centric randomized experiments or natural experiments. In this way, causal conclusions and insights can be provided to product sellers and recommender system practitioners.
\item \xihdr{Algorithms.} Although we only focus on algorithms trained on explicit feedback, it is relatively intuitive to extend the proposed error correlation framework to other pointwise recommendation algorithms. Another direction is to address the marketing bias in pairwise ranking recommendation algorithms, where market fairness metrics and debiasing methods can be further explored to accommodate real-world scenarios.
\end{itemize}

\bibliographystyle{ACM-Reference-Format}
\bibliography{references}

\end{document}